# Interactions of Hydrogen Molecules with Halogen-Containing Diatomics from Ab Initio Calculations: Spherical-Harmonics Representation and Characterization of the Intermolecular Potentials

Alessandra F. Albernaz,[†] Vincenzo Aquilanti,[‡] Patricia R. P. Barreto,*[,§] Concetta Caglioti,[‡] Ana Claudia P. S. Cruz,[§] Gaia Grossi,[‡] Andrea Lombardi,*[,‡] and Federico Palazzetti[‡,#]

[†]Instituto de Física, Universidade de Brasília, CP04455, Brasília, Distrito Federal CEP 70919-970, Brazil

[‡]Dipartimento di Chimica, Biologia e Biotecnologie, Università di Perugia, via Elce di Sotto 8, 06123 Perugia, Italy

[§]Instituto Nacional de Pesquisas Espaciais (INPE)/MCT, Laboratòrio Associado de Plasma (LAP), CP515, São José dos Campos, São Paulo CEP 12247-970, Brazil

[#]Scuola Normale Superiore, Piazza dei Cavalieri 7, 56126, Pisa, Italy.

**Abstract**

For the prototypical diatomic-molecule−diatomic-molecule interactions H2−HX and H2−X2, where X = F, Cl, Br, quantum-chemical ab initio calculations are carried out on grids of the configuration space, which permit a spherical-harmonics representation of the potential energy surfaces (PESs). Dimer geometries are considered for sets of representative leading configurations, and the PESs are analyzed in terms of isotropic and anisotropic contributions. The leading configurations are individuated by selecting a minimal set of mutual orientations of molecules needed to build the spherical-harmonic expansion on geometrical and symmetry grounds. The terms of the PESs corresponding to repulsive and bonding dimer geometries and the averaged isotropic term, for each pair of interacting molecules, are compared with representations in terms of a potential function proposed by Pirani et al. (see Chem. Phys. Lett. 2004, 394, 37−44 and references therein). Connections of the involved parameters with molecular properties provide insight into the nature of the interactions.

## 1. INTRODUCTION

The characterization of the intermolecular potential energy surfaces (PESs) of pairs of simple molecular systems and their compact representation through suitable analytic functions is an essential requirement for applications of classical, semiclassical, and quantum-mechanical approaches to molecular spectroscopy and dynamics. We report (i) the results of quantum-chemical ab initio calculations on a series of interactions between diatomic molecules, specifically, H2−H2, H2−X2, and H2−HX, where X = F, Cl, Br, (ii) their use to assemble full dimensional intermolecular PESs in the framework of a spherical-harmonics expansion, and (iii) the analysis of the features of the interactions, in particular, in terms of their comparison with an intermolecular potential function proposed by Pirani et al.[1,2]

Two alternative attitudes are currently taken to describe weak molecular interactions. These attitudes can be traced back to the permeating view of chemical phenomena from structural and dynamical perspectives. Let us consider the case of an atom interacting with a diatomic molecule to describe the two approaches.

The structural approach is implemented in common quantum-chemical calculations, where one computes energy as a function of a properly defined distance, R, most often the distance between the atom and the center-of-mass of the molecule and an angle θ, typically that formed by the direction of R and the molecular axis. Inspired by open-shell−closed-shell atom−atom interactions, where Σ and Π states are quantum mechanically constructed,[3−5] one would consider a minimal model for a linear or bent triatomic molecular structure including the potential energy profiles, say $V_{\parallel}(R)$ and $V_{\perp}(R)$, corresponding to the collinear (θ = 0) and perpendicular (θ = π/2) approaches as the leading configurations characterizing this system. Spectroscopy methods should provide information on the structure of the complex, and quantum chemistry comparisons will be relevant for an assessment of the nature of the intermolecular bond.

Alternatively, a dynamical viewpoint describes the interaction as a sum of an isotropically averaged contribution, as typically measurable from scattering of an atom with a hot (i.e., fast rotating) diatom and an anisotropic term. In a description of a diatomic molecule as a rigid rotor, the intermolecular potential V(R, θ), where $0 \le \theta \le \pi$, which admits an expansion in spherical harmonics, in this case simply Legendre polynomials $P_\mu(\cos\theta)$ (the orthonormal expansion set with the required completeness properties for convergence for a function defined 0−π range), can be expressed as

$$V(R,\theta) = \sum_{\mu} V_{\mu}(R)\, P_{\mu}(\cos\theta)$$

where μ = 0, 2, ... (the odd terms vanish by symmetry). The "minimal" expansion in this case is obtained by truncation, $V(R, \theta) = V_0(R) + V_2(R) P_2(\cos\theta)$, where $P_2(\cos\theta) = (3\cos^2(\theta) - 1)/2$. Identifying $V_{\parallel} = V(R, 0)$ and $V_{\perp} = V(R, \pi/2)$, one has the relations

$$V_0(R) = \frac{3}{2}(V_{\parallel} - V_{\perp})$$

$$V_2(R) = \frac{1}{3}(V_{\parallel} + V_{\perp})$$

or inversely $V_{\parallel} = V_0 + V_2$ and $V_{\perp} = V_0 - 1/2\ V_2$

Experimentally, the $V_\mu(R)$ coefficients may come for a class of generally simple cases from delicate molecular beam experiments and indirectly from the analysis of transport properties sensitive to molecular anisotropies (see, e.g., refs 6 and 7 and references therein).

The second view illustrated above can be recognized as an archetype of the spherical-harmonics expansion, such as that of refs 8−11, where the mathematically rigorous connections between configurations of selected approach angles and the corresponding expansion moments are established. Historically, a structural view focuses on the V(R, θ) for which appears the minimum leading configuration, more insightful for bound systems (the molecular limit). In contrast, the dynamical view focuses on the isotropic $V_0$ term and is more insightful for weakly bonded systems. A recent example where both views have been compared is for the interactions in the rare-gas−$H_2O_2$ systems.[12,13]

This paper is a continuation of a series in which quantum chemical studies were reported on a variety of combinations of approaching molecules, the spherical-harmonics expansion coefficients correlated to the characterizing leading configurations and the results being interpreted according to progress in our understanding of intermolecular interactions. Here the program is extended to systems containing the simplest diatomic molecule $H_2$ with $H_2$, HF, HCl, and HBr to study the progression to molecules having a halogen as a source of substantial electronegativity.

The coefficients of the expansion terms are usually given in functional forms such as those of Morse and Dunham,[14,15] analytically convenient for the vibrational manifold of the anharmonic oscillator model of diatomics and directly exportable to calculations of the structure and particularly of the dynamics, while insight into the nature of the weak chemical bond comes from the analysis of the isotropic and anisotropic terms. Our analysis exploits the recently developed Pirani formula, which besides the two basic parameters (well depth, $D_e$, and well position, $R_{eq}$) introduces two additional ones, m characterizing the type of long-range of interaction and its "hardness'" β related to the sharpness of the curvature. The Pirani potential function is best suited for reproducing the van der Waals part of the interaction and small anisotropies caused by charge-transfer or permanent dipole or quadrupole moment perturbative contributions. When more pronounced anisotropic effects are present, it can be conveniently combined with explicit electrostatic terms.

The sections of this paper are organized as follows. In the next section, background information on the theoretical method adopted for the calculation of the potential profiles for a number of leading configurations is outlined. Details of the spherical-harmonics expansion method are briefly resumed. The Pirani function is exploited to enlighten the physical meaning of the corresponding parameters. In Section 3, the results of the ab initio calculations are presented and the use of the Pirani potential as a fitting function and its validation in a series of cases is reported. The paper ends with Section 4 reporting conclusions and perspectives.

## 2. POTENTIAL ENERGY SURFACES

In this section we briefly outline the approach followed to build up global intermolecular PESs for the H2−HX and H2−X2 systems, with X = H, F, Cl, Br, based on ab initio calculations of potential energy profiles for a series of leading configurations, defined following a spherical-harmonics expansion of the neutral−neutral molecule interaction. The energy profiles are smoothly glued together by the spherical-harmonics expansion. Then, the use of the Pirani potential function is experimented to fit the isotropic part, exploiting the use of insightful parameters of the interaction, and to obtain efficient simplified models in those cases where the anisotropy due to additional terms coming from electrostatic or charge-transfer contributions is limited to a perturbative effect, to be indirectly enclosed in the Pirani function.

**2.1. Leading Configurations.** The leading configurations can be better defined by considering the spherical-harmonics expansion of the intermolecular PES of two diatomic molecules. The earliest one was introduced and used for the O2−O2 system,16,17 subsequently extended to cover heteronuclear diatomics (a general formulation is given in ref 18). For a given configuration of the two molecules, treated as rigid rotors, at a certain value of the distance R of their centers of mass, the intermolecular interaction energy V depends on a set of three angular coordinates, denoted as θa, θb, ranging from 0 to π, and ϕ, ranging from 0 to 2π, as follows

$$V(R,\theta_a,\theta_b,\phi) = 4\pi \textstyle\sum_{L_a,L_b,L} V^{L_a,L_b,L} Y^{L0}_{L_a,L_b}(\theta_a,\theta_b,\phi), \qquad (1)$$

where La, Lb = 0, 1, 2, ... and | La − Lb | ≤ L ≤ La + Lb and the angular functions YLa,Lb L0 are bipolar spherical harmonics

$$Y^{L0}_{L_a,L_b}(\theta_a,\theta_b,\phi) = \textstyle\sum_m (-)^{L_a-L_b} \begin{pmatrix} L_a & L_b & L \\ m & -m & 0 \end{pmatrix} \times \Upsilon_{L_a,m}(\theta_a,\phi_a)\Upsilon_{L_b,-m}(\theta_b,\phi_b) \qquad (2)$$

where the functions YLa,m and YLb,−m are ordinary spherical harmonics the symbol between large parentheses is a 3 – j symbol19 and it holds the inequality −min(La,Lb) ≤ min(La,Lb). The radial coefficients VLa,Lb,L (R) are the "momenta" of the expansion and, in general, incorporate the radial part of different components of the interaction, such as dispersion induction, and electron overlap plus electrostatic contributions. Figure 1 shows the set of leading configurations that have been chosen for characterizing the interactions of the pair of diatomic molecules considered here along with the corresponding values of the angular coordinates. At any given configuration, the spherical-harmonics in eq 1 assume fixed values and the potential energy, depending on R, is reduced to a linear combination of the corresponding momenta VLa,Lb,L (R). For a value Ri assigned to the distance R, the interaction energy can be estimated here by appropriate ab initio calculations, for each of the leading configurations, and a set of values for cuts of the PES is obtained. The obtained values and the corresponding sets of the VLa,Lb,L (R) radial momenta serve to set up a system of linear equations that can be solved analytically, expressing the radial momenta as a combination of spherical-harmonics. The physical meaning of the isotropic radial term, the V000(R) moment of the expansion, is remarkable: We will want to interpret it in terms of the relative contribution of the size-repulsion, induction, and dispersion attraction terms to the intermolecular interaction. It can also include those additional attractive effects (as those due to charge-transfer components), which may not vanish when averaging over all mutual orientations.

**2.2. Pirani Potential for a Pair of Interacting Centers.** The Lennard-Jones (12−6) is a well-known potential model that describes the weak interactions of a pair of atoms or neutral molecules as a sum of two contributions: a repulsive component of the potential that depends on R−12 where R is the interatomic (or intermolecular) distance and an attractive component that depends on R−6. The interaction potential is given by the following equation

$$V_{LJ}(R) = D_e\left[\left(\frac{R_{eq}}{R}\right)^{12} - 2\left(\frac{R_{eq}}{R}\right)^{6}\right] \qquad (3)$$

where De is the well depth and Req is the equilibrium distance. Although the Lennard-Jones (12−6) model gives a realistic description of the potential around the distance corresponding to the minimum energy of the system, it does not accurately reproduce the interaction both at long and at short distances. The first important attempt to overcome the limits of the Lennard-Jones potential was proposed by Maitland and Smith,20 who modified the VLJ potential as follows

$$V_{MS}(R) = D_e\left[\frac{6}{n(x)-6}\left(\frac{1}{x}\right)^{n(x)} - \frac{n(x)}{n(x)-6}\left(\frac{1}{x}\right)^{6}\right] \qquad (4)$$

In such potential, x is the reduced distance R/Req and n(x) = 13 + g (x − 1), where g is a parameter depending on the specific case and ranging between 2 and 13. A generalization of the Maitland and Smith model is that proposed by Pirani et al.,1,2 where n(x) = β + 4x2 and β is a dimensionless parameter. The authors provided indications for the best values of β. For a broad class of systems β depends on the hardness of the interacting atoms or molecules. More precisely, β in the range 7−10 can be used to describe systems involving atom/ molecules that are highly or not very polarizable and where electrostatic forces, such as ion-permanent dipole and ion−ion interactions, are present at long-range.21 A further parameter m assumes a value 6 for two neutral partners, 4 for ion-induced dipole interactions, 2 for ion-permanent dipole interactions, and 1 for ion−ion systems. The Pirani potential function reads as follows

$$V_P(R) = D_e\left[\frac{m}{n(x)-m}\left(\frac{1}{x}\right)^{n(x)} - \frac{n(x)}{n(x)-m}\left(\frac{1}{x}\right)^{6}\right] \qquad (5)$$

We denote this potential as Pirani's because the previously used acronym ILJ (Improved Lennard-Jones) is misleading, being essentially an improvement of the one by Maitland and Smith, which was a proposed improvement of Lennard- Jones's. It has been demonstrated that differently from the classical Lennard-Jones model the Pirani function correctly reproduces the long-range behavior of the potential for a wide variety of systems.2,22 The function in eq 5 is versatile due to the relationship of the parameters De, Req, m, and β with the molecular polarizability. (See eqs 6 and 7 in Section 2.3 and ref 23 for details of correlation formulas.) For that reason, it can be applied to more complex molecular systems, exploiting the additivity of the polarizabilities in building the interactions as a result of a collection of interacting centers. This particularly interesting feature of such an approach has been applied to triatomic molecules,24−30 small aggregates,31 and the liquid state32−34 and has been widely used in the modeling of the interactions in plasma kinetics applications.35,36

**2.3. Calculation of Potential Energy Profiles and Correlation Analysis of the Parameters.** The intermolecular interaction energy profiles for any given pair of molecules and for each leading configuration (see Figure 1) have been obtained according to the following procedure. For each configuration, kept frozen holding fixed the corresponding values of the angular coordinates, a set of ab initio energies was calculated at the CCSD(T)/aug-cc-pVQZ level of theory, corresponding to 100 different values of the distance R between the two molecular centers of mass, sampled by setting an equally spaced grid extending from approximately 2 to 8 Å. A fitting procedure of the ab initio points for each pair of molecules and for each leading configuration, 6 for the H2−X2 systems and 9 for the H2−XH, was carried out using generalized Rydberg functions of fifth degree,37 and the corresponding depths and distances of the wells were obtained. For each pair of interacting molecules, the energy profiles of all related leading configurations were collected and used to generate the spherical-harmonics expansion, as previously illustrated (Section 2.1) The isotropic part of the interaction V000(R) (see eq 1) was generated by averaging over the corresponding leading configurations, which is equivalent to an integration over a discretized angle domains for the θa, θb, and ϕ angular coordinates. Afterward, the ab initio points were also fitted using the Pirani functional of eq 5, where the parameters m, β, Req, De, are involved. An additional step consisted of a Dunham expansion15 of the rovibrational energies from which the vibrational frequencies, ωe, of the PESs for each leading configuration were obtained.

The parameter β of the Pirani function (see eq 5) is a measure of "hardness" of the molecular pair interaction. For comparison, an estimate of it, denoted as βc, can be obtained from the molecule polarizabilities according to the following formula38

$$\beta_c = 6 + \frac{5}{\alpha_a^{1/3} + \alpha_b^{1/3}} \qquad (6)$$

where αa and αb are the polarizabilities in Å3. This insight into the parameter β makes it worth to work out an expression for the vibrational frequency in terms of the Pirani parameters

$$\omega_e = \sqrt{\frac{D_{eq}(4+\beta)m}{4\pi^2 R_{eq}^2 \mu}} \qquad (7)$$

where μ represents the reduced mass of the system.

As a consequence of the relationships of eqs 6 and 7 for any given intermolecular PES we were able to compare the sets of parameters De, Req, β, m, and ωe of the corresponding Pirani fit to the analogous ones of Rydberg (De and Req) and Dunham (ωe) fitting functions, plus a comparison of the values of β with those of βc.

Because of its relationship to polarizabilities (see eq 6), βc has been chosen as a correlation parameter for Req, De, m, β, and ωe. Linear fitting by the formula

$$P(\beta_c) = a + b\beta_c$$

with P = β, De, Req, ωe, m, and the a and b fitting coefficients have been calculated for some of the leading configurations of Figure 1 for a pair of interacting molecules of types H2-HX and $H_2-X_2$, as detailed in the next section.

Note that the Pirani function of eq 5 is best suited for themodeling of the isotropic part of the interaction, that is, the $V_{000}$ coefficient of eq 1. Nevertheless, in many cases the intermolecular interaction of neutral molecules is characterized by a leading size repulsion plus dispersion attraction contribution perturbed by small additional ones, such as those due to the induction effects associated with permanent dipole or quadrupole moments, charge transfer, or purely electrostatic terms. In such situations, this generally small perturbation can be incorporated into eq 5 by properly adjusting the β or m parameters to take into account their modulation effects on the repulsive wall and the attraction tail. Under such conditions, the Pirani function of the global intermolecular interaction can be adopted to represent the full interaction, also for specific cuts (in particular, those corresponding to the leading configurations) of the PES.

**3. RESULTS**

According to the procedure outlined in Section 2.3, for a set of leading configurations (see Figure 1) we calculated the PES corresponding to a number of some significantly repulsive potential energy profiles (L1 and H), to others (L, T2, T3) showing wells, and also to the isotropic terms $V_{000}(R)$. These are presented in Figures 2−5, where the ab initio energy points as a function of the distance R (see Section 2.1) are shown along with the corresponding Rydberg and Pirani fitting functions.

The ab initio points have been fitted by using the Pirani function potential by varying the parameters β and m (see eq 5). Figure 2 shows for the systems $H_2-HX$ (X = F, Cl, Br) the repulsive configuration L1, while Figure 3 shows the nonmonotonic potential profile of the T3 configuration and the isotropic term of the spherical-harmonics expansion for the same systems. Figure 4 shows for the systems $H_2-X_2$ (X = H, F) the nonmonotonic energy profiles of the L and T2 leading configurations and the isotropic term. Figure 5 shows for the systems $H_2-X_2$ (X = Cl, Br) the repulsive potential energy profile of the H leading configuration, the nonmonotonic energy profiles of the H and T2 configurations, and the isotropic term.

In nearly all considered cases (except for some repulsive curves), the adequacy of the Pirani fitting function in reproducing the energy profile is excellent and comparable to that of the fifth degree Rydberg functions[37] (see Section 2.3). The equilibrium energy and distance increase according to the atomic number of the atoms X (= H, F, Cl, Br) involved (see also Table 1), and a similar behavior is found for the repulsive wall of the potential that shifts to larger distances for the heavier atoms of the series, as expected due to steric hindrance. The intermolecular range where the potential well extends is indeed mainly ascribed to the dispersion forces that increase with the polarizability of the molecule, which is higher when larger atoms are involved (see Table 2). The interaction potential of the L configuration stands out for its highly repulsive contribution.

The dispersion forces play an important role in the stability of the system in the $H_2-F_2$ case (see Figure 4). For $H_2-Cl_2$ and $H_2-Br_2$ (see Figure 5) the electrostatic quadrupole− quadrupole components increase their role (the quadrupole moments have the same sign) so that the potential, in the H leading configuration, is exclusively repulsive. The intermediate panels show the interaction potential related to the minimum energy configuration that corresponds to the T2 configuration for the $H_2-X_2$ systems (see Figures 4 and 5) and to the T3 configuration for the $H_2-HX$ systems (see Figure 3). The restrained steric hindrance and the presence of dispersion forces, combined with small electrostatic effects, make the $H_2-F_2$ system the most stable one, with the highest absolute value of minimum energy and the shortest equilibrium distance, while for $H_2-Cl_2$ and $H_2-Br_2$ (see Figure 5) the equilibrium distance is placed at highest values, while the well depth decreases, as expected. The trend of the interaction potential for the $H_2-H_2$ systems (see Figure 4) does not change too much in the considered configurations and in the isotropic term of the spherical-harmonics expansion because of the small molecular dimensions and the low anisotropy of the polarizability. The isotropic component of the intermolecular potential for the $H_2-HX$ systems (X = F, Cl, Br) compared with that of the $H_2-H_2$ system shows a deeper energy well being strongly affected by the increase in dispersion/induction attraction in the $H_2-HX$ systems.

A resumé of the absolute minima and equilibrium distances, calculated with the use of correlation formulas illustrated in ref 23 and by CCSD(T) calculations, is reported in Table 2 for the interacting systems considered here. Results by correlation formula are shown with or without the specific induction contribution attractive terms due to the interaction between permanent and induced dipole moments and obtained by the relationship given in ref 23, where the dipole moment and the polarizability of the interacting molecules are used.

As anticipated in Section 2.3, the Pirani potential can be used as a fitting function, where the physically meaningful parameters, m and β, the equilibrium energy, De, and th equilibrium distance, Req, can be adjusted to reproduce the potential energy profiles corresponding to any given leading configuration. Also, the vibrational frequency, ωe, can be included in the list of parameters to be monitored for which the correlation is evaluated because it can be easily expressed as function of β (see eq 7) and m. It must be remarked that the Pirani function, as formulated in eq 5, is tailored to model the isotropic part of the intermolecular interaction established between a pair of molecules and also defined cuts of the full PES; it is basically controlled by the short-range size repulsion and the dispersion/induction attraction contributions.39 On the contrary, the function can incorporate perturbative effects due to, for example, charge transfer or electrostatic components b slightly adjusting the parameters β and m.21

To investigate the suitability of the Pirani function as a model extended to cases including additional interaction components, in the following we analyze the linear correlation between any single of the previously mentioned parameters and the calculated hardness βc (see eq 6), as detailed in the previous section (see eq 8), for any pair of interacting molecules H2−HX and H2−X2. The calculated hardness βc takes decreasing values in passing from the F to the Br atom in the interacting pair of both H2−HX and H2−X2 types.

Table 1 shows the calculated polarizabilities of the molecules considered in the present work and the values of βc for any given pair obtained by eq 6. Figures 6−8 show the linear fit of the Pirani parameters, m and β, the equilibrium energy, De, the equilibrium distance, Req, and the frequency, ωe, versus the increasing βc trend in the H2−HX and H2−X2 pair interactions, with X = H, F, Cl, Br, as obtained from eq 6 (see Table 1). The calculated potential energy profiles that have been considered for the fitting procedure are of three kinds: the V000 isotropic terms, the nonmonotonic Vmin, and the fully repulsive Vr. The values of the Pirani parameters versus βc, that have been obtained in the present analysis, are collected in Tables 3 and 4. It can be seen that the closeness of the adjusted values of the Pirani potential parameters to their typical values indicated in Section 2.2 is maximum when the isotropic terms V000 of any of the interacting molecular pairs are fitted. This is not surprising, even in cases where significant electrostatic contributions are present (e.g., the H2−HF pair), because these contributions are canceled by the spherical average that defines the V000 term of the spherical-harmonics expansion (see Section 2.1). The Pirani function here adopted is in fact best suited for reproducing the isotropic part of the interaction.

On the contrary, the presence of significant additional positive electrostatic contributions to the interaction (as typical, for example, of molecules interacting already at long range with a large permanent-dipole−permanent-quadrupole moment) makes the potential repulsive at any R; see, for instance, the L1 configuration for H2−HF (Figure 2, upper panel). This causes the values of the Pirani potential parameters to deviate from their expected values. In these cases, the use of the Piranifunction only for modeling purposes, for example, to obtain tentative PESs to be subsequently refined by ab initio calculations, is not recommended as such, because most of the original physical meaning of the parameters is lost. Instead, in many intermediate cases, where small additional contributions to the isotropic one are present, as, for example, charge-transfer effects or electrostatic contributions (charge distributions with nonzero dipole or quadrupole moments), a slight adjustment of the β and m parameters can lead to efficiently incorporate the correction in the Pirani function. This possibility has been extensively exploited in several recent applications (see refs 21, 31, 33, and 39 and references therein and refs 24, 30, and 32).

## 4. CONCLUSIONS

We have investigated the intermolecular interactions between pairs of diatomic molecules of the type H2−HX and H2−X2 for X = F, Cl, Br. The aim of the paper was that of showing how an integrated structural and dynamic view can lead, with the help of selected quantum-chemical calculations, to full dimensional analytic intermolecular PESs. Also, by use of an especiall parametrized function (the Pirani potential), an analysis has been carried out of the main features of the interactions in terms of physically motivated parameters and contributions. This work also suggests that the use of the Pirani potential can be useful to build up force fields in analytic form, suitable for calculations aimed at the evaluation of structural and dynamical properties of the matter.

## ACKNOWLEDGMENTS

Discussions and encouragement from Professor Fernando Pirani are gratefully acknowledged. A.L. acknowledges financial support from the Dipartimento di Chimica, Biologia e Biotecnologie dell'Universitá di Perugia (FRB grant), from MIUR PRIN 2010- 2011 (contract 2010ERFKXL_002) and from "Fondazione Cassa Risparmio Perugia (Codice Progetto: 2015.0331.021 Ricerca Scientifica e Tecnologica)". F.P., A.L., and V.A. acknowledge the Italian Ministry for Education, University and Research, MIUR for financial support through SIR 2014 Scientific Independence for Young Researchers (RBSI14U3VF) and FIRB 2013 Futuro in Ricerca (RBFR132WSM_003). V.A. thanks CAPES for the appointment as Professor Visitante Especial at Instituto de Fìsica, Universidade Federal de Bahia, Salvador (Brazil).

**REFERENCES**

(1) Pirani, F.; Albertí, M.; Castro, A.; Moix Teixidor, M.; Cappelletti, D. The Atom-bond Pairwise Additive Representation for the Intermolecular Potential Energy Surfaces. Chem. Phys. Lett. 2004, 394, 37−44.

(2) Pirani, F.; Brizi, S.; Roncaratti, L. F.; Casavecchia, P.; Cappelletti, D.; Vecchiocattivi, F. Beyond the Lennard-Jones Model: a Simple and Accurate Potential Function Probed by High Resolution Scattering Data Useful for Molecular Dynamics Simulations. Phys. Chem. Chem. Phys. 2008, 10, 5489−5503.

(3) Aquilanti, V.; Grossi, G. Angular Momentum Coupling Schemes in the Quantum Mechanical Treatment of P-state Atom Collisions. J. Chem. Phys. 1980, 73, 1165−1172.

(4) Aquilanti, V.; Beneventi, L.; Grossi, G.; Vecchiocattivi, F. Coupling Schemes for Atom-diatom Interactions and an Adiabatical Decoupling Treatment for Rotational Temperature Effects on Glory Scattering. J. Chem. Phys. 1988, 89, 751−761.

(5) Aquilanti, V.; Liuti, G.; Pirani, F.; Vecchiocattivi, F. Orientational and Spin Orbital Dependence of Interatomic Forces. J. Chem. Soc., Faraday Trans. 2 1989, 85, 955−964.

(6) Aquilanti, V.; Bartolomei, M.; Cappelletti, D.; Carmona-Novillo, E.; Pirani, F. Dimers of the Major Components of the Atmosphere: Realistic Potential Energy Surfaces and Quantum Mechanical Prediction of Spectral Features. Phys. Chem. Chem. Phys. 2001, 3, 3891−3894.

(7) Aquilanti, V.; Ascenzi, D.; Bartolomei, M.; Cappelletti, D.; Cavalli, S.; de Castro Vitores, M.; Pirani, F. Molecular Beam Scattering of Aligned Oxygen Molecules. The Nature of the Bond in the O2-O2 Dimer. J. Am. Chem. Soc. 1999, 121, 10794−10802.

(8) Palazzetti, F.; Munusamy, E.; Lombardi, A.; Grossi, G.; Aquilanti, V. Spherical and Hyperspherical Representation of Potential Energy Surfaces for Intermolecular Interactions. Int. J. Quantum Chem. 2011, 111, 318−332.

(9) Barreto, P. R. P.; Albernaz, A. F.; Palazzetti, F.; Lombardi, A.; Grossi, G.; Aquilanti, V. Hyperspherical Representation of Potential Energy Surfaces: Intermolecular Interactions in Tetra-atomic and Penta-atomic Systems. Phys. Scr. 2011, 84, 028111.

(10) Barreto, P. R. B.; Albernaz, A. F.; Capobianco, F.; Palazzetti, A.; Lombardi, A.; Grossi, G.; Aquilanti, V. Potential Energy Surfaces for Interactions of H2O with H2, N2 and O2: a Hyperspherical Harmonics Representation, and a Minimal Model for the H2O-rare-gas-atom Systems. Comput. Theor. Chem. 2012, 990, 53−61.

(11) Barreto, P. R. P.; Vilela, A. F. A.; Lombardi, A.; Maciel, G. S.; Palazzetti, F.; Aquilanti, V. The Hydrogen Peroxide-Rare Gas Systems: Quantum Chemical Calculations and Hyperspherical Harmonic Representation of the Potential Energy Surface for Atom-Floppy Molecule Interactions. J. Phys. Chem. A 2007, 111, 12754−12762.

(12) Maciel, G. S.; Barreto, P. R. P.; Palazzetti, F.; Lombardi, A.; Aquilanti, V. A Quantum Chemical Study of H2S2: Intramolecular Torsional Mode and Intermolecular Interactions with Rare Gases. J. Chem. Phys. 2008, 129, 164302.

(13) Roncaratti, L. F.; Leal, L. A.; Pirani, F.; Aquilanti, V.; e Silva, G. M.; Gargano, R. Chirality of Weakly Bound Complexes: The Potential Energy Surfaces for the Hydrogen-peroxide-noble-gas Interactions. J. Chem. Phys. 2014, 141, 134309.

(14) Morse, P. M. Phys. Rev. 1929, 34, 57.

(15) Dunham, J. L. The Energy Levels of a Rotating Vibrator. Phys. Rev. 1932, 41, 721.

(16) Aquilanti, V.; Ascenzi, D.; Bartolomei, M.; Cappelletti, D.; Cavalli, S.; de Castro Vitores, M.; Pirani, F. Quantum Interference Scattering of Aligned Molecules: Bonding in O4 and Role of Spin Coupling. Phys. Rev. Lett. 1999, 82, 69−72.

(17) Aquilanti, V.; Ascenzi, D.; Bartolomei, M.; Cappelletti, D.; Cavalli, S.; de Castro Vitores, M.; Pirani, F. Molecular Beam Scattering of Aligned Oxygen Molecules. The nature of the Bond in the O2-O2 Dimer. J. Am. Chem. Soc. 1999, 121, 10794−10802.

(18) Barreto, P.; Cruz, A. C.; Barreto, R.; Palazzetti, F.; Albernaz, A.; Lombardi, A.; Maciel, G.; Aquilanti, V. The Spherical-Harmonics Representation for the Interaction Between Diatomic Molecules: The General Case and Applications to CO-CO and CO-HF. 2015, in preparation.

(19) Carmona-Novillo, E.; Pirani, F.; Aquilanti, V. Quantum Dynamics of Clusters on Experimental Potential Energy Surfaces: Triplet and Quintet O2-O2 Surfaces and Dimers of para-N2 with ortho and para-N2 and with O2. Int. J. Quantum Chem. 2004, 99, 616−627.

(20) Maitland, G. C.; Smith, E. B. A Simplified Representation of Intermolecular Potential Energy. Chem. Phys. Lett. 1973, 22, 443.

(21) Albertí, M.; Aguilar, A.; Cappelletti, D.; Laganà, A.; Pirani, F. On the Development of an Effective Model Potential to Describe Water Interactions in Neutral and Ionic Clusters. Int. J. Mass Spectrom. 2009, 280, 50−56.

(22) Lombardi, A.; Palazzetti, F. A Comparison of Interatomic Potentials for Rare Gas Nanoaggregates. J. Mol. Struct.: THEOCHEM 2008, 852, 22−29.

(23) Barreto, P. R. P.; Palazzetti, F.; Grossi, G.; Lombardi, A.; Maciel, G. S.; Vilela, A. F. A. Range and Strength of Intermolecular Forces for van der Waals Complexes of the Type H2XnRg, with X = O, S and n = 1, 2. Int. J. Quantum Chem. 2010, 110, 777−786.

(24) Bartolomei, M.; Pirani, F.; Laganà, A.; Lombardi, A. A Full Dimensional Grid-empowered Simulation of the CO2 + CO2 Processes. J. Comput. Chem. 2012, 33, 1806−1819.

(25) Lombardi, A.; Faginas Lago, N.; Laganà, A.; Pirani, F.; Falcinelli, S. A Bond-Bond Portable Approach to Intermolecular Interactions: Simulations for N-methylacetamide and Carbon Dioxide Dimers. In ICCSA 2012, Part I, LNCS; Springer Verlag: Berlin, 2012; Vol. 7333, pp 387−400.

(26) Falcinelli, F.; Rosi, M.; Candori, P.; Vecchiocattivi, F.; Bartocci, A.; Lombardi, A.; Faginas Lago, N.; Pirani, F. Modeling the Intermolecular Interactions and Characterization of the Dynamics of Collisional Autoionization Processes. In ICCSA 2013, Part I, LNCS; Springer Verlag: Berlin, 2013; Vol. 7971, pp 69−83.

(27) Lombardi, A.; Laganà, A.; Pirani, F.; Palazzetti, F.; Faginas-Lago, N. Carbon Oxides in Gas Flows and Earth and Planetary Atmospheres: State-to-State Simulations of Energy Transfer and Dissociation Reactions. In Computational Science and Its Applications- ICCSA 2013; Springer Verlag, 2013; Vol. 7972, pp 17−31.

(28) Lombardi, A.; Faginas-Lago, N.; Pacifici, L.; Costantini, A. Modeling of Energy Transfer From Vibrationally Excited CO2 Molecules: Cross Sections and Probabilities for Kinetic Modeling of Atmospheres, Flows, and Plasmas. J. Phys. Chem. A 2013, 117, 11430−

11440.

(29) Lombardi, A.; Faginas-Lago, N.; Pacifici, L.; Grossi, G. Energy Transfer upon Collision of Selectively Excited CO2 Molecules: State to- state Cross Sections and Probabilities for Modeling of Atmospheres and Gaseous Flows. J. Chem. Phys. 2015, 143, 034307.

(30) Lombardi, A.; Bartolomei, M.; Pirani, F.; Laganà, A. Energy Transfer Dynamics and Kinetics of Elementary Processes in Planetary Atmosphere Environments: Selectivity Control by the Anisotropy of the Interaction. J. Comput. Chem. 2016, n/a.

(31) Albertí, M.; Amat, A.; Aguilar, A.; Huarte-Larrańaga, F.; Lucas, J. M.; Pirani, F. A Molecular Dynamics Study of the Evolution from the Formation of the C6H6-(H2O)n Small Aggregates to the C6F6 formation. Theor. Chem. Acc. 2015, 134, 61-1−61-12.

(32) Faginas-Lago, N.; Lombardi, A.; Albertí, M.; Grossi, G. Accurate Analytic Intermolecular Potential for the Simulation of Na+ and K+ Ion Hydration in liquid Water. J. Mol. Liq. 2015, 204, 192−197.

(33) Albertí, M.; Amat, A.; Farrera, L. I.; Pirani, F. From the (NH3)2−5 Clusters to Liquid Ammonia: Molecular Dynamics Simulations Using NVE and NpT Ensembles. J. Mol. Liq. 2015, 212,

307−315.

(34) Faginas-Lago, N.; Albertí, M.; Laganà, A.; Lombardi, A. Ion- Water Cluster Molecular Dynamics Using a Semiempirical Intermolecular Potential. In ICCSA 2015, Part II, LNCS; Springer Verlag,

2015; Vol. 9156, pp 355−370.

(35) Celiberto, R.; Armenise, I.; Cacciatore, M.; Capitelli, M.; Esposito, F.; Gamallo, P.; Janev, R.; Laganà, A.; Laporta, V.; Laricchiuta, A.; et al. Atomic and Molecular Data for Spacecraft Reentry

Plasmas. Plasma Sources Sci. Technol. 2016, in press.

(36) Laganà, A.; Lombardi, A.; Pirani, F.; Gamallo, P.; Sayós, R.; Armenise, I.; Cacciatore, M.; Esposito, F.; Rutigliano, M. Molecular Physics of Elementary Processes Relevant to Hypersonics: Atom-Molecule, Molecule-Molecule and Atoms-Surface Processes. Open Plasma Phys. J. 2014, 7, 48−59.

(37) Rydberg, R. Graphische Darstellung einiger bandenspektroskopischer Ergebnisse. Eur. Phys. J. A 1931, 73, 376−385.

(38) Capitelli, M.; Cappelletti, D.; Colonna, G.; Gorse, C.; Laricchiuta, A.; Liuti, G.; Longo, S.; Pirani, F. On the Possibility of Using Model Potentials for Collision Integral Calculations of Interest for Planetary Atmospheres. Chem. Phys. 2007, 338, 62−68.

(39) Cappelletti, D.; Bartocci, A.; Grandinetti, F.; Falcinelli, S.; Belpassi, L.; Tarantelli, F.; Pirani, F. Experimental Evidence of Chemical Components in the Bonding of Helium and Neon with Neutral Molecules. Chem. - Eur. J. 2015, 21, 6234−6240.

(40) Anderson, D. T.; Schuder, M.; Nesbitt, D. J. Large-amplitude Motion in Highly Quantum Clusters: High-resolution Infrared Absorption Studies of Jet-cooled H2HCl and H2DCl. Chem. Phys. 1998, 239, 253.

(41) Diep, P.; Johnson, J. K. An Accurate H2-H2 Interaction Potential from First Principles. J. Chem. Phys. 2000, 112, 4465.

(42) Scheer, A. M.; Mozejko, P.; Gallup, G. A.; Burrow, P. D. Total Dissociative Electron Attachment Cross Sections of Selected Aminoacids. J. Chem. Phys. 2007, 126, 174301.

**Figure 1.**

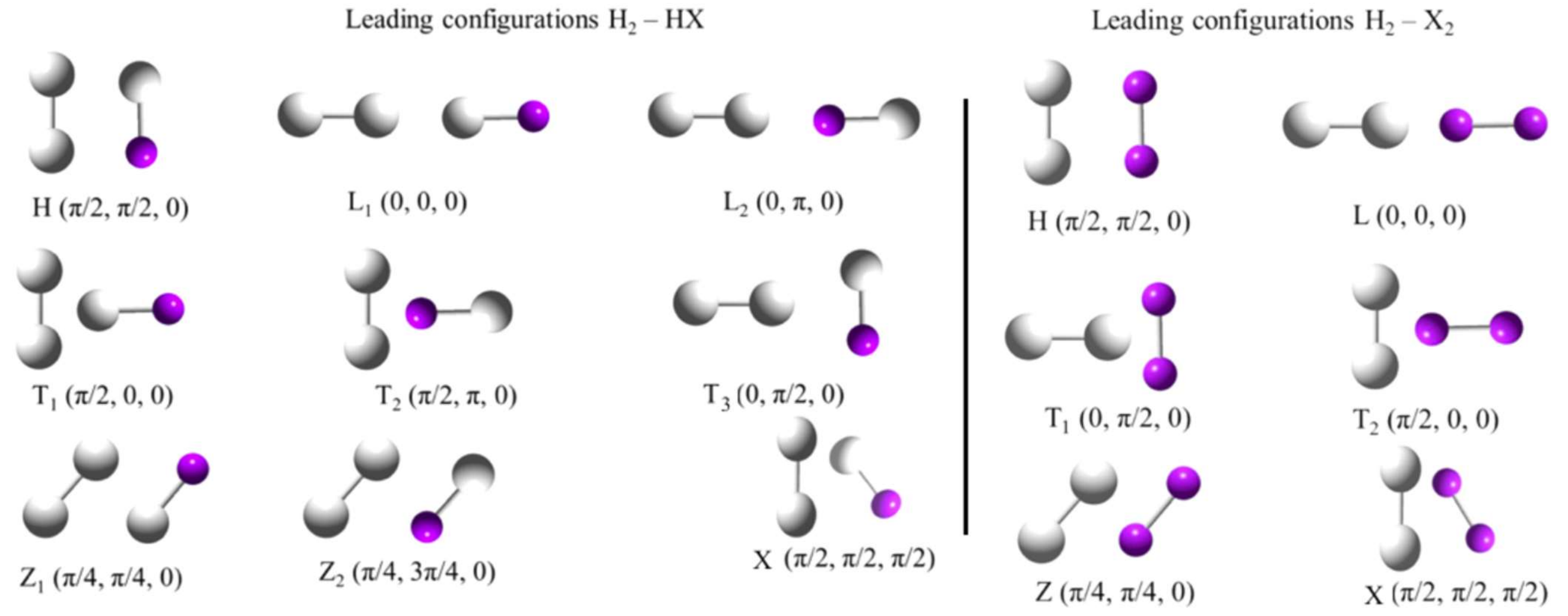

Leading configurations H2 – HX
H (π/2, π/2, 0)
L1 (0, 0, 0)
L2 (0, π, 0)
T1 (π/2, 0, 0)
T2 (π/2, π, 0)
T3 (0, π/2, 0)
Z1 (π/4, π/4, 0)
Z2 (π/4, 3π/4, 0)
X (π/2, π/2, π/2)
Leading configurations H2 – X2
H (π/2, π/2, 0)
L (0, 0, 0)
T1 (0, π/2, 0)
T2 (π/2, 0, 0)
Z (π/4, π/4, 0)
X (π/2, π/2, π/2)

**Figure 2.**

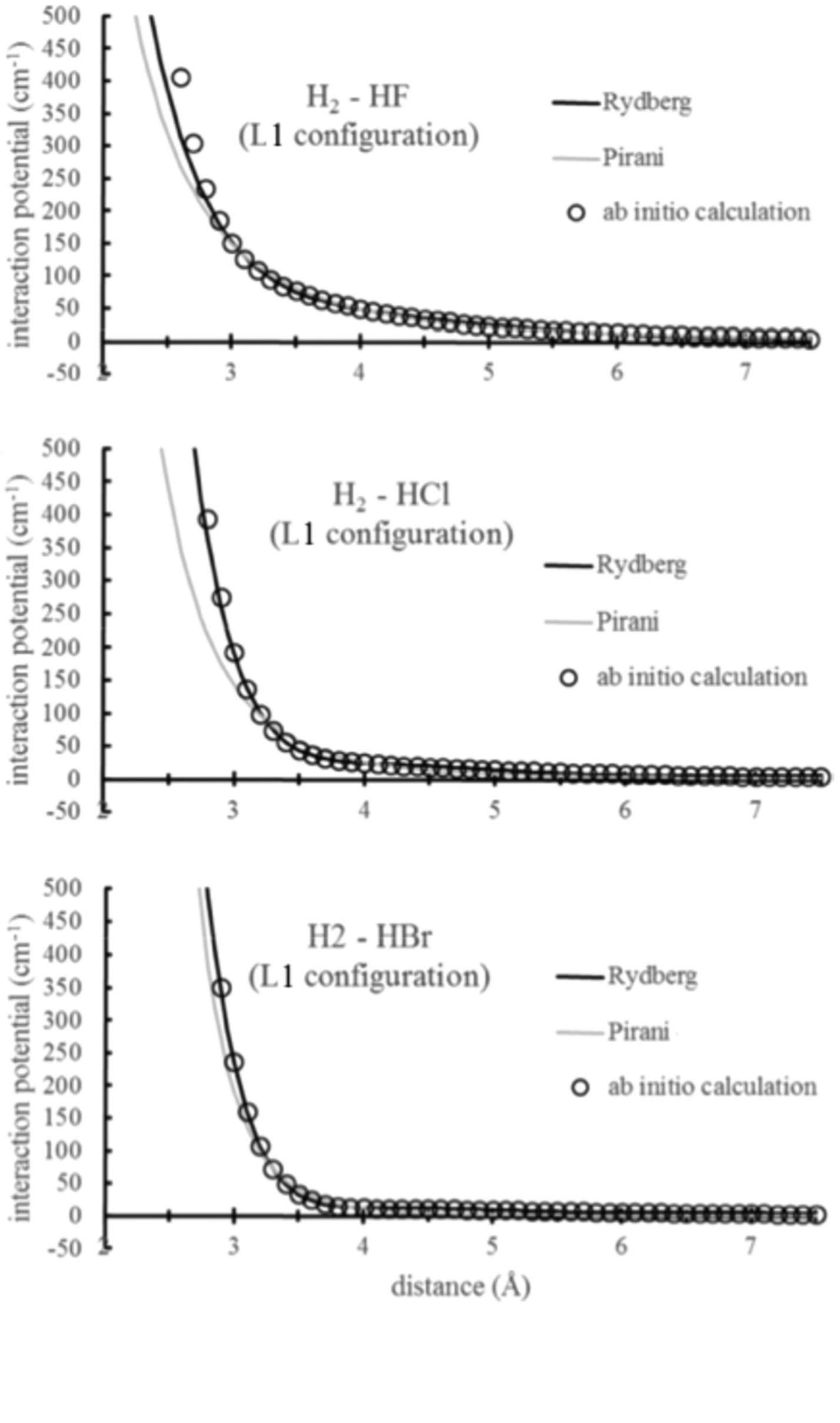

**Figure 3.**

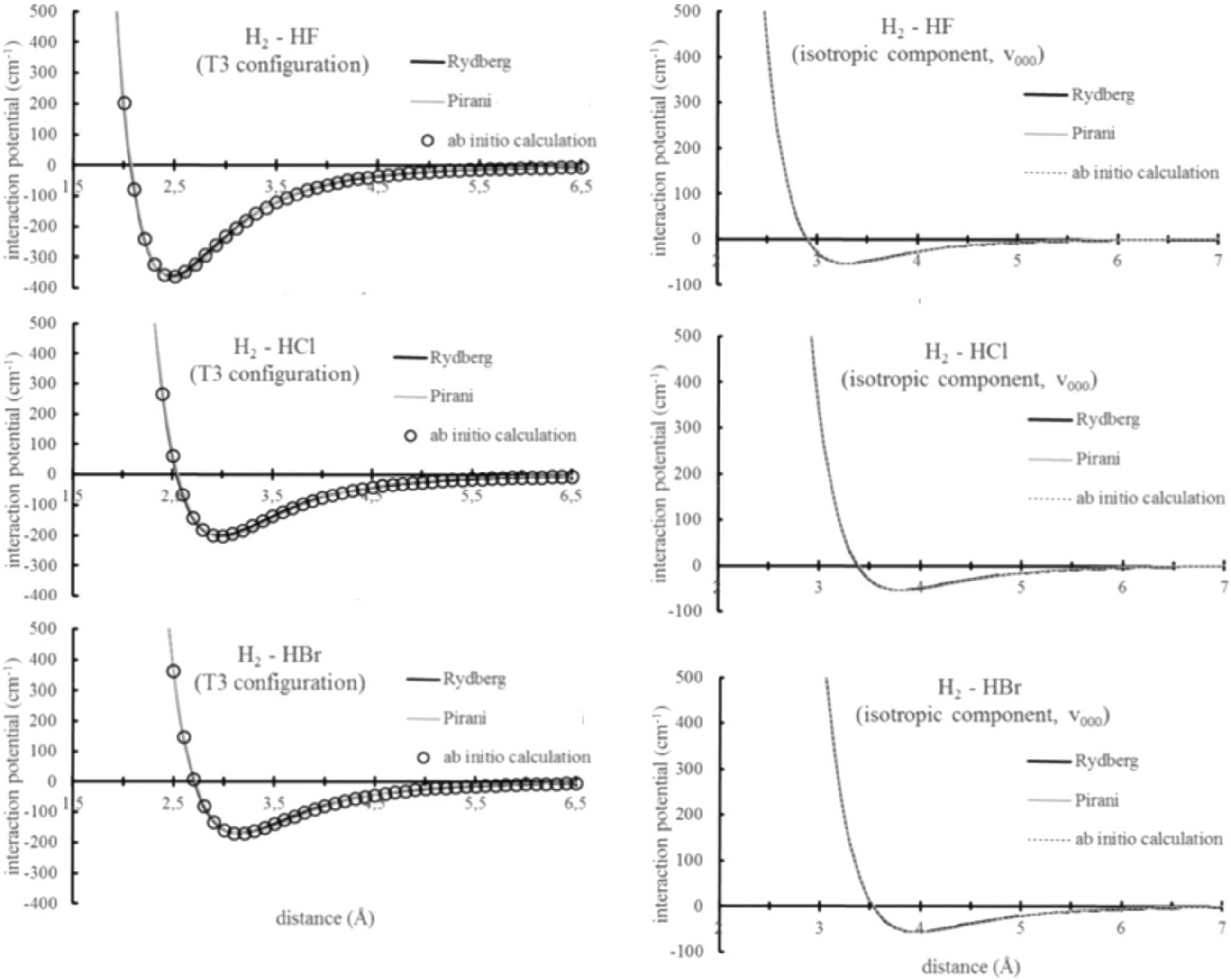

**Figure 4.**

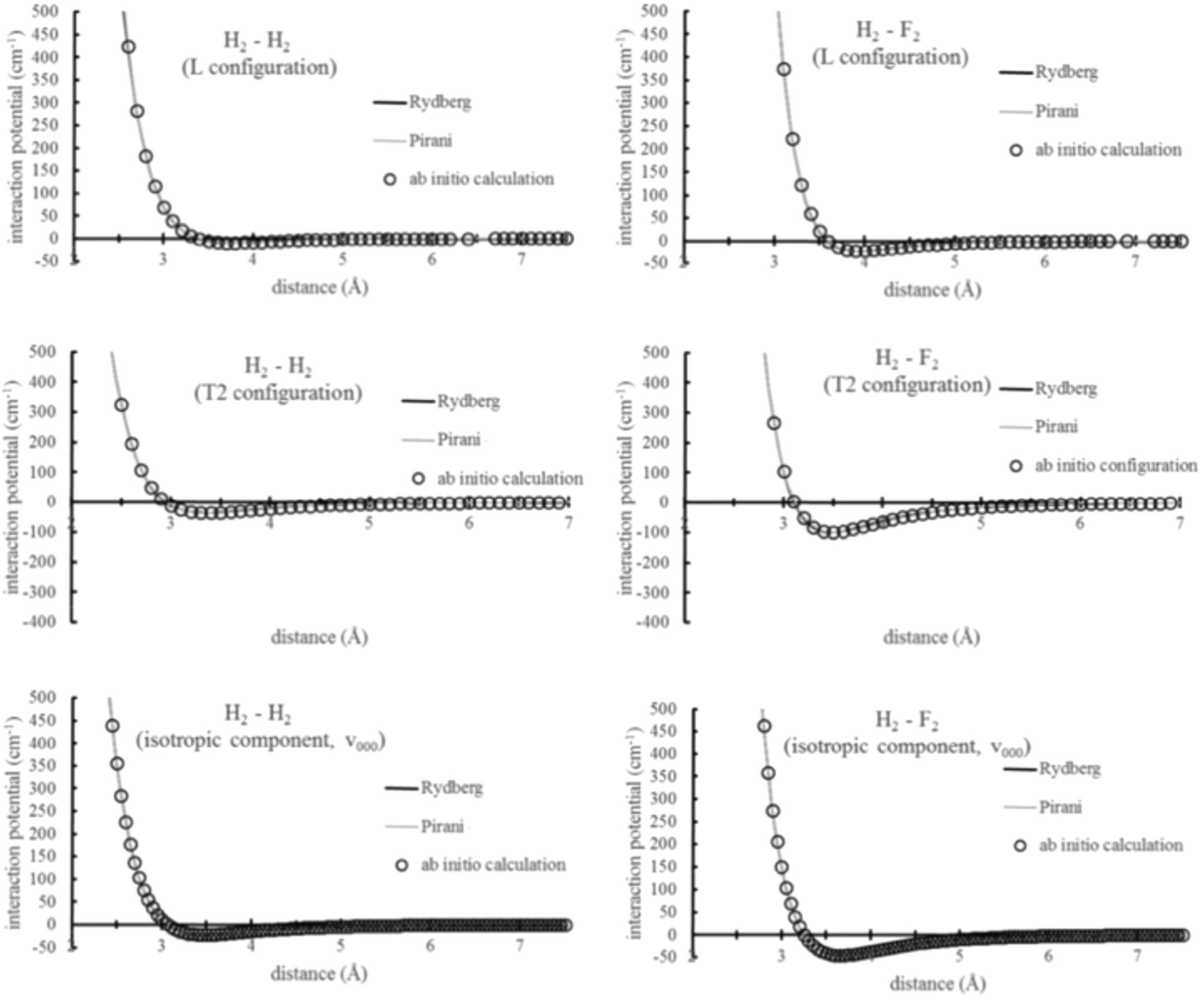

**Figure 5.**

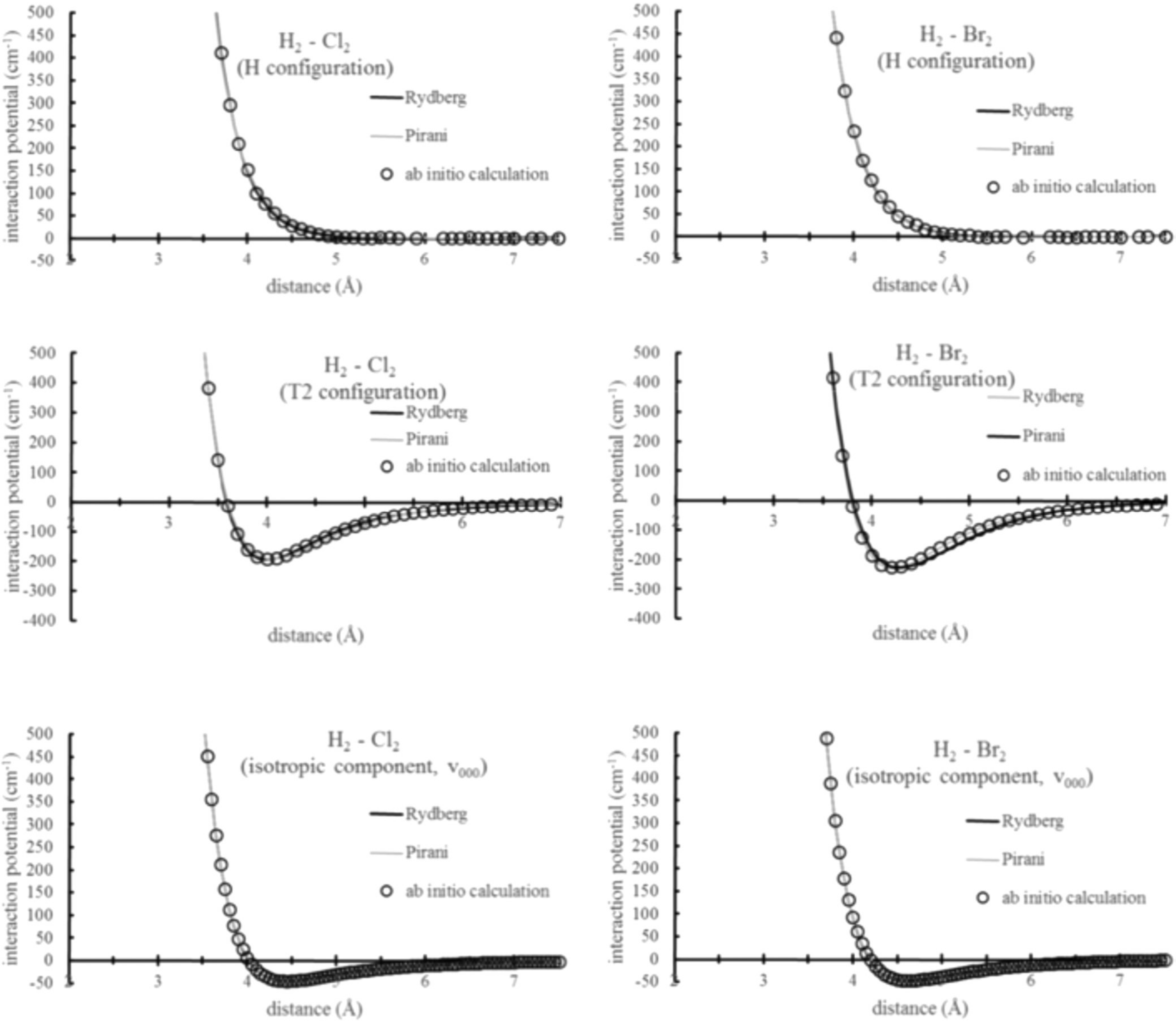

**Figure 6.**

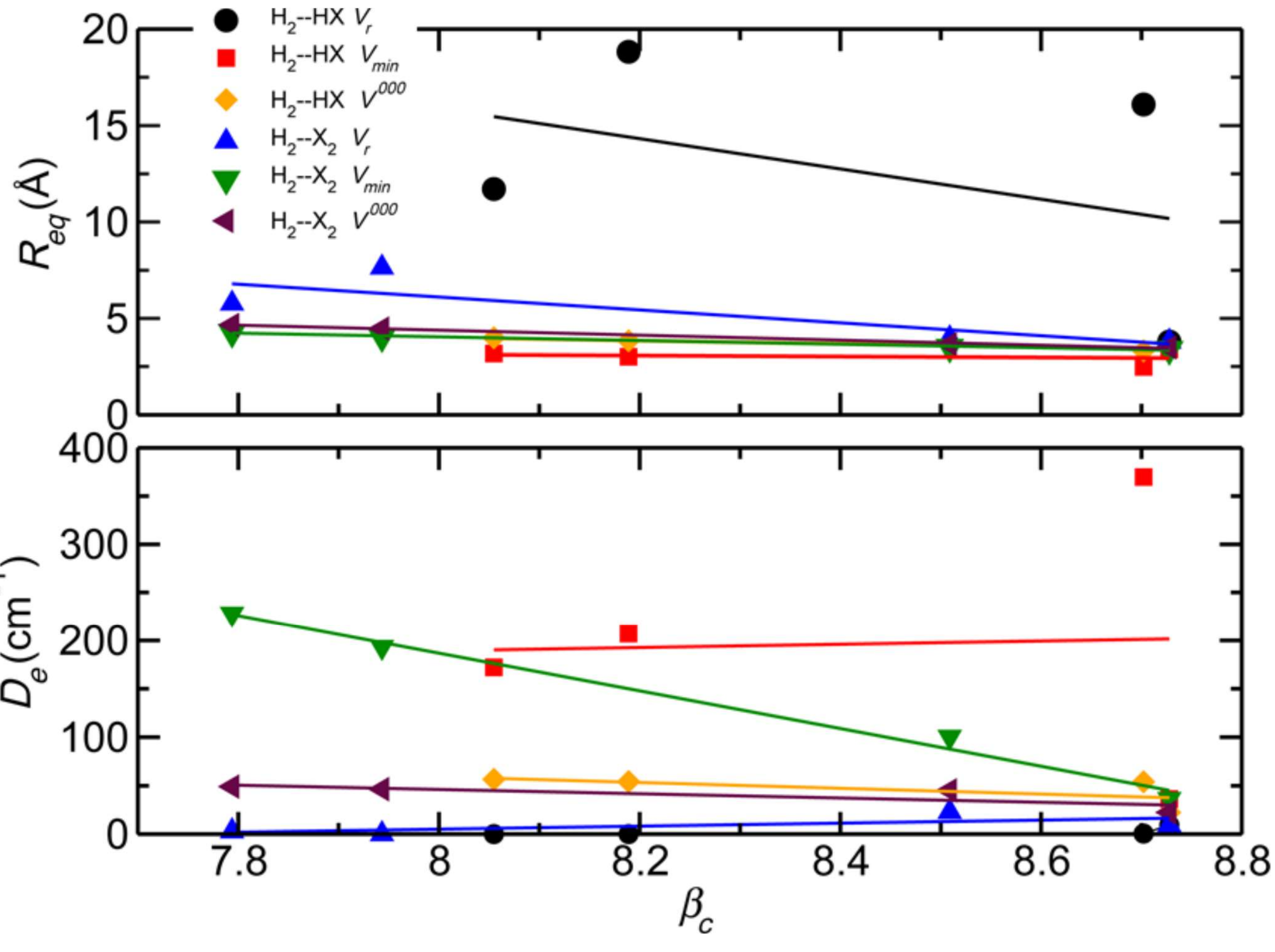

**Figure 7.**

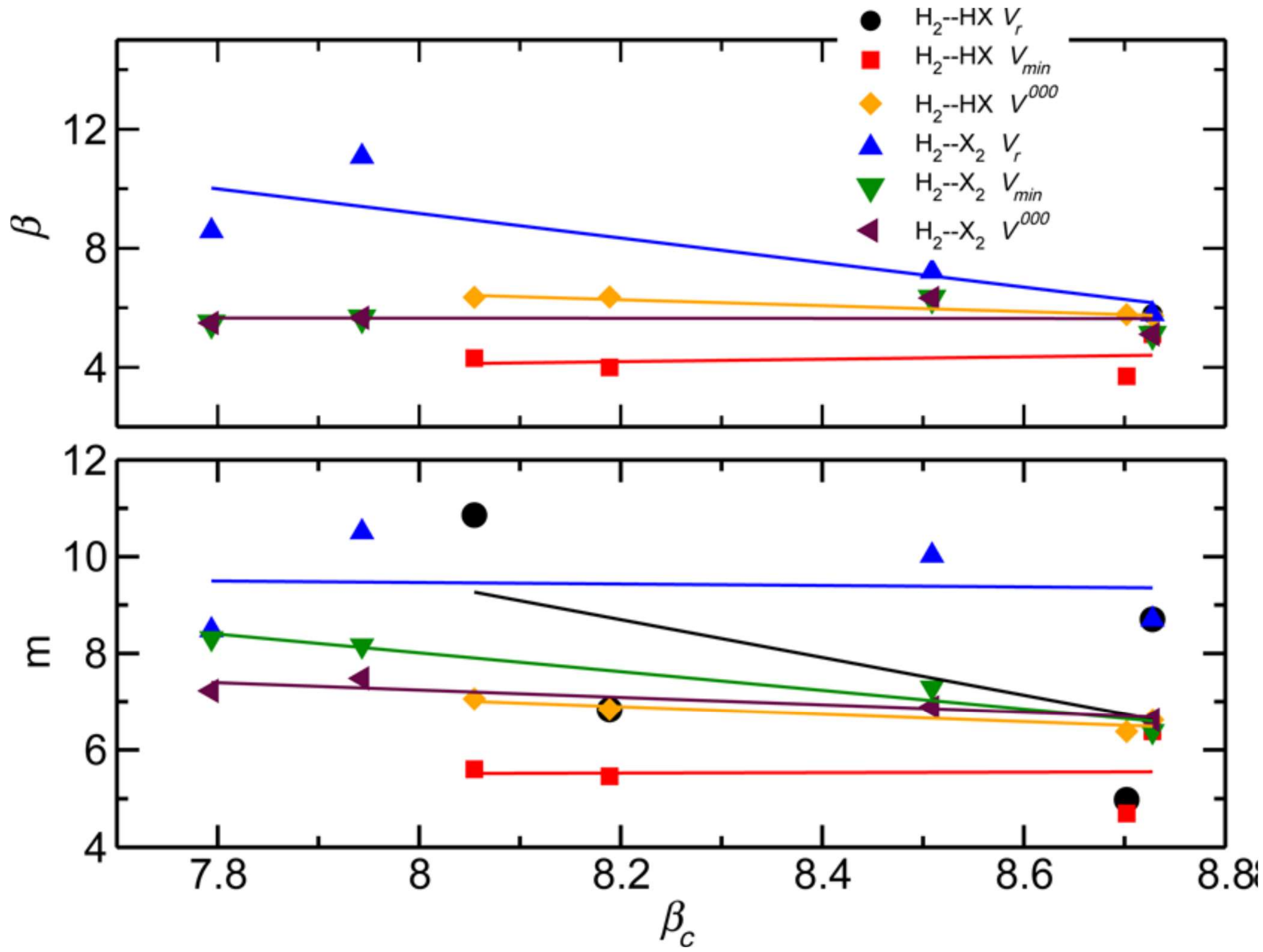

**Figure 8.**

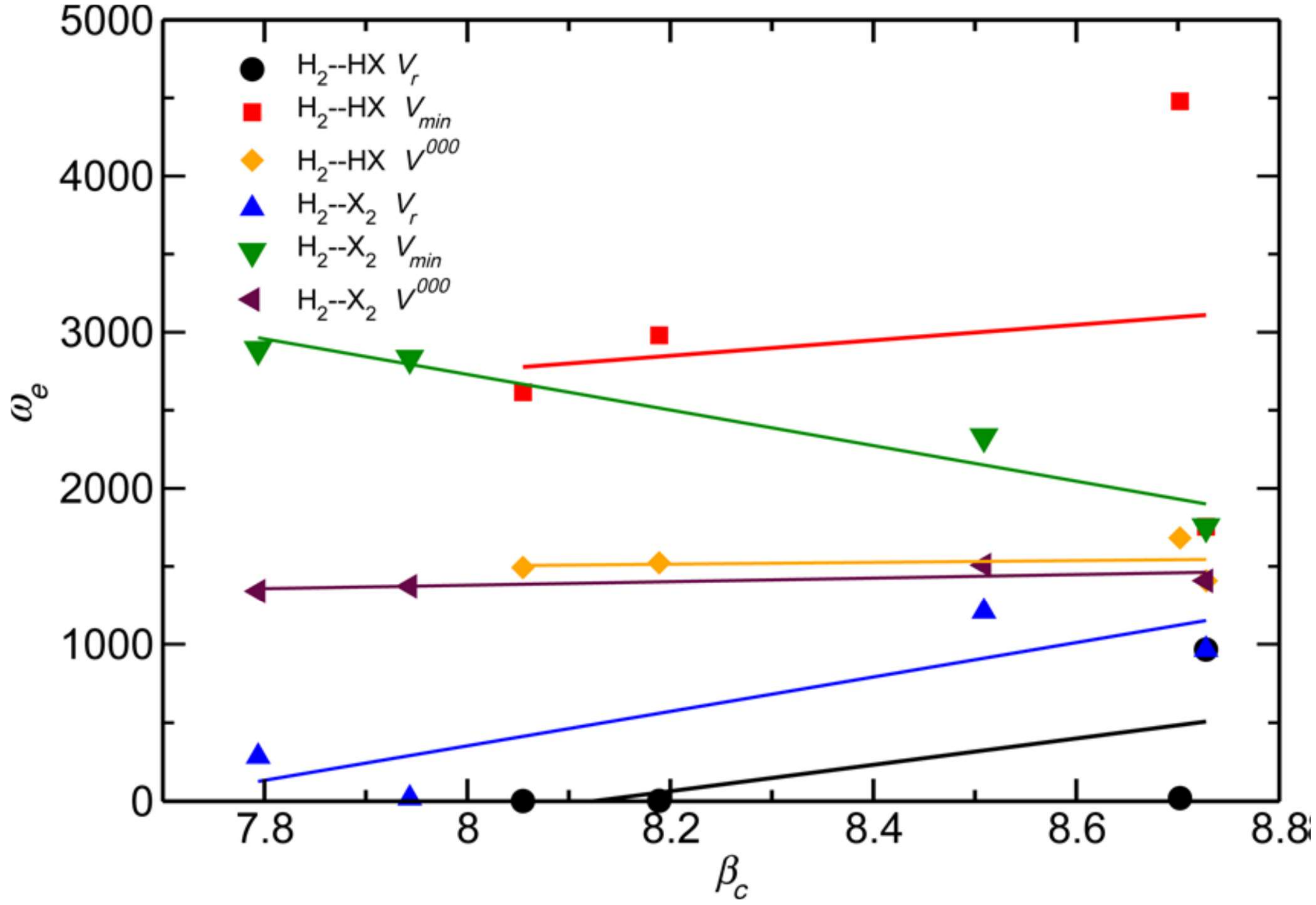

**Figure Captions.**

**Figure 1.** Sketch of the leading configurations adopted to characterize the intermolecular interactions of the H2−X2 and H2−HX systems. The corresponding values of the θa, θb, and ϕ coordinates are indicated.17,18

**Figure 2.** Ab initio potential energy profiles and Rydberg anPirani fitting for the H2−HX systems (X = F, Cl, Br) in the L1 configuration.

**Figure 3.** Ab initio potential energy profiles, Rydberg and Pirani fitting for the H2−HX systems (X = F, Cl, Br) in the T3 configuration, and isotropic components of the spherical-harmonics expansions.

**Figure 4.** Ab initio potential energy profiles, Rydberg and Pirani fitting of the H2−H2 and H2−HF systems in the T2 configuration, and isotropic components of the spherical-harmonics expansions.

**Figure 5.** Ab initio potential energy profiles, Rydberg and Pirani fitting of the H2−Cl2 and H2−Br2 systems in the H and T2 configurations, and isotropic components of the spherical-harmonics expansions.

**Figure 6.** Correlation between the Pirani function parameters, Deq and Req, and the calculated hardness, βc.

**Figure 7.** Correlation between the Pirani function parameters, β and m, and the calculated hardness, βc.

**Figure 8.** Correlation between the dimer vibrational frequencies, ωe, and the calculated hardness, βc.

**Table 1.** Calculated Polarizabilities (Present Work) in Å3 of HX and X2 Molecules (X = H, F, Cl, Br) and Values of βc for H2 − HX and H2 − X2 Systems

| | H2 | HF | HCl | HBr | F2 | Cl2 | Br2 |
|---|---|---|---|---|---|---|---|
| $\alpha$ | 0.770 | 0.814 | 2.558 | 3.490 | 1.247 | 4.545 | 6.547 |

$\beta$ c for interacting pairs H2 − HX(X2)

| | H2 | HF | HCl | HBr | F2 | Cl2 | Br2 |
|---|---|---|---|---|---|---|---|
| H2 | 8.728 | 8.702 | 8.189 | 8.055 | 8.509 | 7.943 | 7.794 |

**Table 2.** Well Depth and Position (in units of cm−1 and Å, respectively) for the Most Stable Configuration of HX and X2 Molecules (X = H, F, Cl, Br) As Obtained with and without Induction Contribution23 and by CCSD(T)/aug-cc-pVQZ Sets.

| | with induction[23] | | no induction[23] | | CCSD(T) | | ref | |
|---|---|---|---|---|---|---|---|---|
| | $R_{eq}$ | $D_e$ | $R_{eq}$ | $D_e$ | $R_{eq}$ | $D_e$ | | |
| $H_2$ | 3.404 | 27.994 | 3.404 | 27.994 | 3.474 | 22.033 | 3.434[40] | 25.124[41] |
| | | | | | | | 3.48200[41] | 24.117[40] |
| $F_2$ | 3.535 | 44.689 | 3.535 | 44.690 | 3.659 | 45.475 | | |
| $Cl_2$ | 4.036 | 58.715 | 4.036 | 58.720 | 4.446 | 46.100 | | |
| $Br_2$ | 4.224 | 61.438 | 4.224 | 61.438 | 4.627 | 48.381 | | |
| HF | 3.391 | 42.466 | 3.418 | 37.332 | 3.296 | 53.059 | | |
| HCl | 3.780 | 53.263 | 3.785 | 52.277 | 3.809 | 3.990 | 53.398[42] | 45.000[42] |
| $F_2$ | 3.535 | 44.689 | 3.535 | 44.689 | 3.659 | 45.475 | | |
| HBr | 3.913 | 58.564 | 3.915 | 58.086 | 3.964 | 55.901 | | |

**Table 3.** Fitting Parameters of the Pirani Function of the Fitted V000, Vr, and Vmin for the H2−HX Interactions (X = H, F, Cl, Br) and Comparison with the Corresponding Rydberg Fit Parameters

| | Pirani function | | | | | Rydberg | | |
|---|---|---|---|---|---|---|---|---|
| | $m$ | β | $R_{eq}$ (Å) | $D_e$ (cm$^{-1}$) | $\omega_e$ (s$^{-1}$) | $R_{eq}$ (Å) | $D_e$ (cm$^{-1}$) | $\omega_e$ (s$^{-1}$) |
| | | | | $V_r$ | | | | |
| $H_2$ | 8.703 | 5.787 | 3.797 | 9.373 | 965.793 | 3.752 | 9.276 | 1037.491 |
| HF | 4.977 | 0.175 | 16.093 | 0.524 | 19.741 | 8.298 | 0.000 | 0.060 |
| HCl | 6.833 | 0.174 | 18.826 | 0.012 | 2.934 | 7.063 | 4.744 | 0.008 |
| HBr | 10.860 | 0.435 | 11.710 | 0.001 | 1.785 | 6.671 | 0.000 | 0.040 |
| | | | | $V_{min}$ | | | | |
| $H_2$ | 6.383 | 5.109 | 3.397 | 36.277 | 1754.302 | 3.366 | 35.526 | 1802.037 |
| HF | 4.696 | 3.698 | 2.485 | 369.705 | 4478.539 | 2.464 | 360.605 | 4478.562 |
| HCl | 5.458 | 3.994 | 3.007 | 207.047 | 2980.080 | 2.971 | 202.824 | 3076.729 |
| HBr | 5.603 | 4.305 | 3.179 | 172.089 | 2614.848 | 3.144 | 171.157 | 2729.504 |
| | | | | $V^{000}$ | | | | |
| $H_2$ | 6.621 | 5.730 | 3.498 | 22.361 | 1408.036 | 3.474 | 22.033 | 1442.436 |
| HF | 6.378 | 5.770 | 3.319 | 53.973 | 1682.448 | 3.296 | 53.059 | 1720.224 |
| HCl | 6.838 | 6.362 | 3.837 | 54.186 | 1522.593 | 3.809 | 53.405 | 1575.504 |
| HBr | 7.058 | 6.349 | 3.995 | 56.439 | 1492.785 | 3.964 | 55.895 | 1556.026 |

**Table 4.** Fitting Parameters of the Pirani Function of the Fitted V000, Vr, and Vmin for the H2−X2 Interactions (X = H, F, Cl, Br) and Comparison with the Corresponding Rydberg Fit Parameters

| | Pirani function | | | | | Rydberg | | |
|---|---|---|---|---|---|---|---|---|
| | $m$ | β | $R_{eq}$ (Å) | $D_e$ (cm$^{-1}$) | $\omega_e$ (s$^{-1}$) | $R_{eq}$ (Å) | $D_e$ (cm$^{-1}$) | $\omega_e$ (s$^{-1}$) |
| $H_2$ | 8.703 | 5.787 | 3.797 | 9.373 | 965.793 | 3.752 | 9.276 | 1037.491 |
| $F_2$ | 10.023 | 7.212 | 3.967 | 23.074 | 1208.761 | 3.937 | 22.811 | 1266.508 |
| $Cl_2$ | 10.510 | 11.075 | 7.638 | 0.015 | 19.083 | 6.022 | 0.000 | 162.928 |
| $Br_2$ | 8.480 | 8.578 | 5.769 | 2.971 | 284.911 | 5.952 | 3.525 | 259.732 |
| | | | | $V_{min}$ | | | | |
| $H_2$ | 6.383 | 5.109 | 3.397 | 36.277 | 1754.302 | 3.366 | 35.526 | 1802.037 |
| $F_2$ | 7.293 | 6.326 | 3.511 | 100.340 | 2330.906 | 3.484 | 100.688 | 2458.959 |
| $Cl_2$ | 8.160 | 5.663 | 4.054 | 193.158 | 2833.052 | 4.008 | 201.946 | 3116.261 |
| $Br_2$ | 8.315 | 5.486 | 4.285 | 227.895 | 2889.271 | 4.219 | 231.824 | 3271.803 |
| | | | | $V^{000}$ | | | | |
| $H_2$ | 6.621 | 5.730 | 3.498 | 22.361 | 1408.036 | 3.474 | 22.033 | 1442.436 |
| $F_2$ | 6.885 | 7.040 | 3.660 | 45.186 | 1507.635 | 3.659 | 45.462 | 1530.417 |
| $Cl_2$ | 7.483 | 8.545 | 4.477 | 46.389 | 1371.883 | 4.447 | 46.067 | 1427.812 |
| $Br_2$ | 7.225 | 8.949 | 4.664 | 49.045 | 1340.903 | 4.627 | 48.387 | 1413.479 |